\documentclass{aa}  

\usepackage{color}
\usepackage{graphicx}
\usepackage{txfonts}
\usepackage{natbib}
\usepackage{url}
\usepackage{xspace}
\bibpunct{(}{)}{;}{a}{}{,}

\newcommand{\teff}{$T_{\rm eff}$}
\newcommand{\gta}{\lower 0.5ex\hbox{$ \buildrel>\over\sim\ $}}
\newcommand{\lta}{\lower 0.5ex\hbox{$ \buildrel<\over\sim\ $}}
\newcommand{\nhe} {$N$({\rm He})/$N$({\rm H})}
\newcommand{\msun}{$M_{\rm \odot}$}

\newcommand{\omcen}{\object{$\omega$ Cen}}

 \usepackage{lineno}
\usepackage{ulem}

\begin{document}

   \title{Just how hot are the $\omega$ Cen extreme horizontal branch pulsators?\thanks{Based on observations (proposal GO-13707) with the NASA/ESA
Hubble Space Telescope, obtained at the Space Telescope
Science Institute, which is operated by the Association of
Universities for Research in Astronomy, Inc., under NASA
contract NAS5-26666.}}

   \author{M. Latour\inst{1}, S. K. Randall\inst{2}, P. Chayer\inst{3}, 
           G. Fontaine\inst{4}, 
           A. Calamida\inst{5}, J. Ely\inst{3}, T. M. Brown\inst{3}, and W. Landsman\inst{6}
          }

   \institute{ Dr. Karl Remeis-Observatory \& ECAP, Astronomical Institute,
               Friedrich-Alexander University Erlangen-N\"{u}rnberg, 
Sternwartstr. 7,
               96049 Bamberg, Germany, \email{marilyn.latour@fau.de}
         \and ESO, Karl-Schwarzschild-Str. 2, 85748 Garching bei M\"{u}nchen, 
Germany 
         \and Space Telescope Science Institute, 3700 San Martin Drive, 
Baltimore, MD 21218, USA
         \and D\'epartement de Physique, Universit\'e de Montr\'eal, Succ. 
Centre-Ville, C.P. 6128, Montr\'eal, QC H3C 3J7, Canada
           \and National Optical Astronomy Observatory - AURA, 950 North Cherry 
Avenue, Tucson, AZ 85719, USA 
             \and Adnet Systems, NASA/GSFC Code 553, Greenbelt, MD 20771, USA}

   \date{Received 25 November 2015; accepted 23 February 2017}

  \abstract
{ Past studies based on optical spectroscopy 
suggest that the five \omcen\ 
pulsators form a rather homogeneous group of hydrogen-rich subdwarf O stars 
with effective temperatures of around 50 000 K. This places the stars 
below the red edge of the theoretical instability strip in the log $g$ $-$ 
\teff\ diagram, where no pulsation modes are predicted to be excited.}
   {Our goal is to determine whether this temperature discrepancy is real, or whether the stars' 
effective temperatures were simply underestimated.}
   {We present a spectral analysis of two rapidly pulsating extreme horizontal 
branch (EHB) stars found in \omcen. We obtained \textit{Hubble Space Telescope}/COS UV spectra of two $\omega$ Cen pulsators, 
V1 and V5, and used the ionisation equilibrium of UV metallic lines to better 
constrain their effective temperatures. As a by-product we also obtained FUV lightcurves of the two pulsators. }
   {Using the relative strength of the \ion{N}{iv} and \ion{N}{v} lines as a 
temperature indicator yields \teff\ values close to 60 000 K, significantly hotter
 than the temperatures previously derived. 
From the FUV light curves we were able to confirm the main pulsation periods known from optical data. 
 }
   {With the UV spectra indicating higher effective temperatures than previously assumed, the sdO stars would
   now be found within the predicted instability strip. Such higher temperatures also provide consistent 
   spectroscopic masses for both the cool and hot EHB stars of our previously studied sample.  }

\authorrunning{M. Latour et al.}  


   \maketitle
%

\section{Introduction}

Hot subdwarf stars populate the blue (thus hot) part of the horizontal branch 
(HB), which is often called the extreme horizontal branch (EHB, 
\citealt{heb08}). Both the HB and EHB are associated with the helium-core 
burning phase of stellar evolution. The peculiarity of the EHB stars is that 
their hydrogen envelope is not massive enough (M \textless 0.02 \msun) to 
sustain significant hydrogen-shell burning. Indeed, hot subdwarf stars have lost 
most of their hydrogen envelope prior to the start of helium-core burning. These 
stars are found in the Galactic field population as well as in several globular 
clusters. An extensive review of these peculiar stars' many properties and the 
current knowledge about them can be found in \citet{heb16}. 

Rapidly pulsating hot subdwarf stars (also known as V361 Hya stars) have been 
known among the field population for 
almost two decades now, since the serendipitious discovery of the first 
pulsating subdwarf B-type (sdB) stars \citep{kil97,koen97}. 
Since then, the number of known V361 Hya stars has increased to more than 50 \citep{ost10}. 
These H-rich pulsating sdBs show multi-periodic luminosity variations, with 
periods of the order of 100$-$200 s and they are found in a well defined 
instability strip between $\approx$29~000 and 36~000 K. Their variability is due to 
pressure ($p$) modes excited by the $\kappa$-mechanism which was found to be 
driven by an increased opacity of iron, and iron-like elements, in the 
sub-photospheric layers of the star \citep{char97}. Radiative levitation is a 
key ingredient in maintaining a sufficient amount of iron in the driving region. 
However some additional diffusion mechanisms, such as mass loss and turbulence \citep{hu11},
 can interact with radiative 
levitation, effectively killing the necessary 
conditions for driving $p$-modes. Indeed, the instability 
strip is far from being pure, with a fraction of pulsators less than about 10\% 
\citep{bil02,ost10}.
Besides these H-rich sdBs, rapid oscillations were also found in SDSS 
J160043.6+074802.9, a hotter subdwarf O-type (sdO) star \citep{wou06}. Despite 
having a high effective temperature ($\approx$68 000 K) as well as a slightly 
enriched helium content (log \nhe~= $-$0.65), the variability of the star 
is thought to arise through the same $\kappa$-mechanism that drives pulsations in sdBs 
\citep{font08,lat11}.

For over a decade, pulsating hot subdwarfs were known only among the field 
population. When rapid oscillations (P $\approx$84-124 s) were first discovered in 
EHB stars in the globular cluster \omcen\ \citep{ran09}, it was assumed that 
these constituted the globular cluster counterparts to the rapid sdB pulsators 
in the field. However, an optical spectroscopic survey at the VLT
revealed that the five known \omcen\ pulsators are in fact He-poor sdO stars 
with
effective temperatures estimated between 48 000-54 000 K \citep{ran11,ran16}. 
This was, and still is, highly intriguing, since the only sdO pulsator currently 
known among the field population is significantly hotter. 
Field counterparts to the \omcen~ pulsators have yet to be found, although
systematic searches have been done \citep{rod07,john14}. 

Apart from \omcen, NGC 2808 is the only other cluster known to host rapid EHB 
pulsators \citep{bro13}. The six known pulsators were found by means of far UV 
photometry with the \textit{Hubble Space Telescope} (\textit{HST}). Low resolution STIS spectra 
were obtained for half of them only, allowing to roughly constrain their 
temperature and atmospheric helium abundance. So far, the NGC 2808 pulsators do 
not appear similar to the ones in \omcen, neither in terms of atmospheric 
parameters nor pulsational properties. 

That being said, to our current knowledge the five \omcen~pulsators, form a 
unique instability strip. This strip has no equivalent, neither in the field, nor in NGC 
2808. A detailed description of the EHB instability strip in \omcen~has recently 
been published by \citet{ran16}. By comparing the position of the stars in the 
log $g$-\teff~diagram with predictions from seismic models, they found the 
pulsators to lay in a region where no pulsation modes are predicted to be 
excited (see their Fig. 11). This is in marked contrast with the pulsating sdB (and the 
one known sdO) stars in the field, for which the driving of pulsations is well 
predicted by the same seismic models.

One explanation discussed in \citet{ran16} is that the temperatures derived from 
optical spectroscopy might underestimate the true effective temperatures of the 
sdO pulsators. 
This phenomenon has been reported in a few sdO stars for which both optical and 
UV spectroscopy are available \citep{fontm08,rauch10,lat15,dix16}. It is related to 
the so-called Balmer line problem: for these stars it is not possible to 
simultaneously reproduce all Balmer lines using the same atmospheric model, the 
higher lines in the series needing a higher \teff\ to be accurately reproduced 
than the lower ones \citep{nap93}. As a consequence, the temperatures derived 
from optical spectra can be misleading and are usually underestimated. 
This problem can be solved to some degree by including metals in the model 
atmospheres used to fit the optical spectra \citep{gia10,rauch14,lat15}. Such 
models yield better fits and higher temperatures. However, this fine-tuning of 
the models to the observed spectra requires good quality optical spectra, since 
the Balmer line problem can be much more subtle in lower quality data. An 
alternative method to estimate temperatures of hot stars is to use the ionization 
equilibrium of metallic species. By fitting metal lines originating from 
different ionization stages of a same element, one can estimate the effective 
temperature \citep{rauch07,fontm08}. This is usually best done for hot stars in 
the UV range, where the strongest metal lines are found. 

Given the faintness of the \omcen\ pulsators (B $\approx$18 mag) and their position 
in a rather crowded field, the quality of optical spectra that can be obtained is 
limited. This is why we turned towards the UV range to provide us with an 
independent temperature determination. We obtained  
$HST$ spectra for two of the sdO pulsators, V1 and V5, with the Cosmic Origin 
Spectrograph (COS). The two stars were chosen as they lie at the cool and hot 
end of the observational instability strip. This paper presents the result of 
our efforts in providing a better estimate of the effective temperature of the 
\omcen\ pulsators and determining whether or not the instability strip 
discrepancy is real.

\section{Mass distribution of the spectroscopic sample}

\begin{figure}
\begin{center}
\includegraphics[height=8cm]{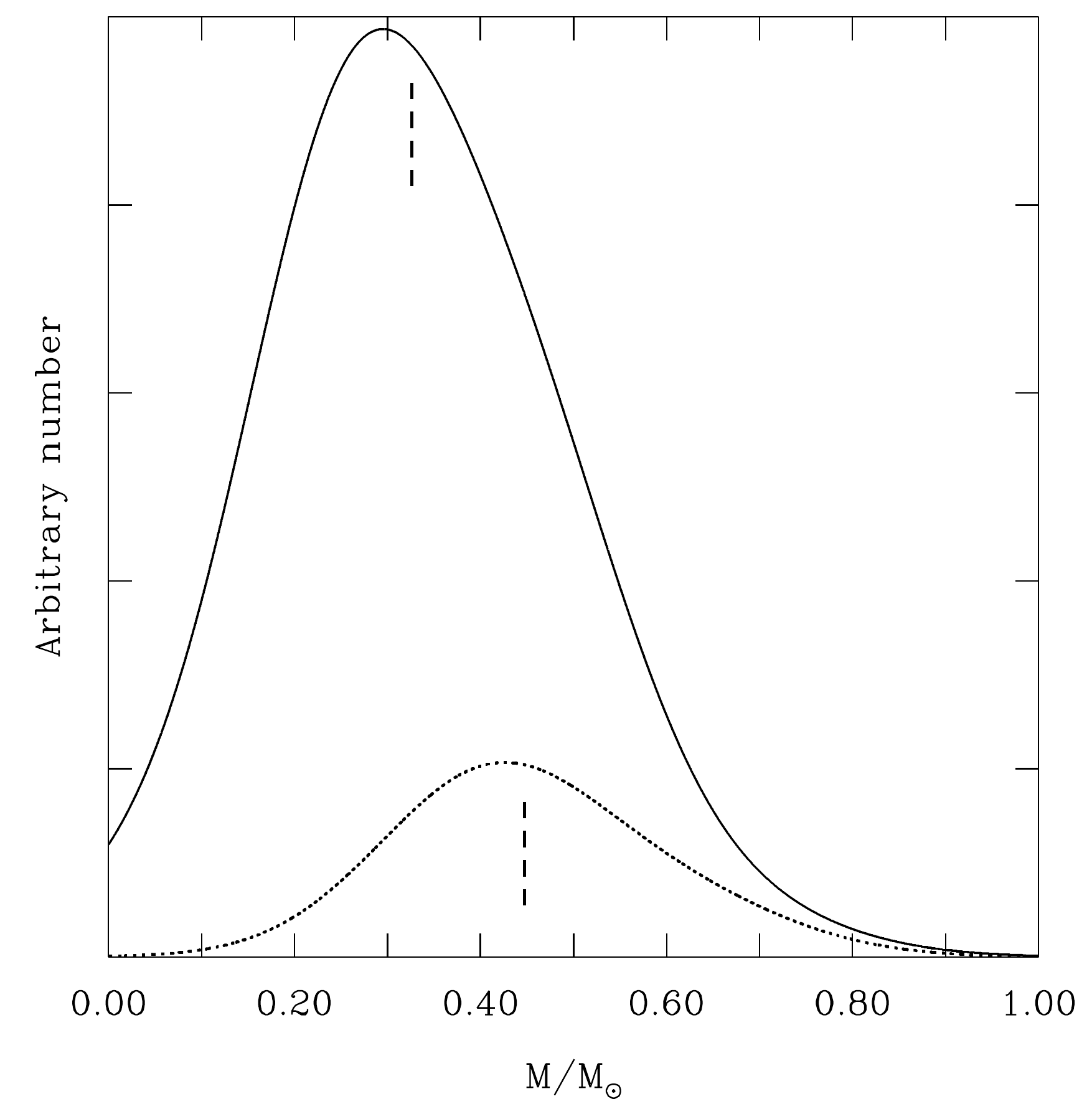}
\caption{Mass distribution of the 32 coolest objects of the sample (solid) and 
equivalent for the subsample of the six hottest stars (dotted).
The mean mass of each distribution (0.331 and 0.452 \msun) is indicated with a 
dash line.}
\label{mass}
\end{center}
\end{figure}

We used the atmospheric parameters determined for 38 EHB stars in \omcen\ 
\citep{lat14} to derive their mass distribution. These 38 stars are the "clean" 
subset described in \citet{ran16} that do not show signs of pollution by a 
cooler star. 
In view of the expected large
uncertainty associated with each individual determination, this is mainly done
with a statistical point of view in mind. At the outset, we
have access to $HST$ Advanced Camera for Surveys (ACS) or
2.2 m MPG/ESO telescope Wide Field Imager (WFI) photometry giving
apparent $B$ magnitudes for all of the 38 stars. Using the
distance modulus of $\omega$ Cen derived in 
\citet{delp06} and \citet{bra16}, $B - M_B$ = 13.70, a reddening index of 
$E(B-V)$ = 0.11
from \citet{cala05}, and a standard Seaton relation, $A_V = 3.20E(B-V)$ \citep{sea79},
we first computed the absolute magnitude $M_B$ of each target
object. 
In a second step, we calculated the theoretical value of $M_B$ from a
synthetic spectrum characterizing each star (\teff, log $g$, and helium 
abundance derived), assuming different given masses. Then parabolic interpolation was
then used to infer the mass of the model with a theoretical absolute
$B$ magnitude that would match the observed value. The resulting masses, as well 
as the atmospheric parameters of the sample, ordered by increasing \teff\ 
are reported in Table \ref{param}. 

\begin{figure*}
\begin{center}
\includegraphics[width=19cm]{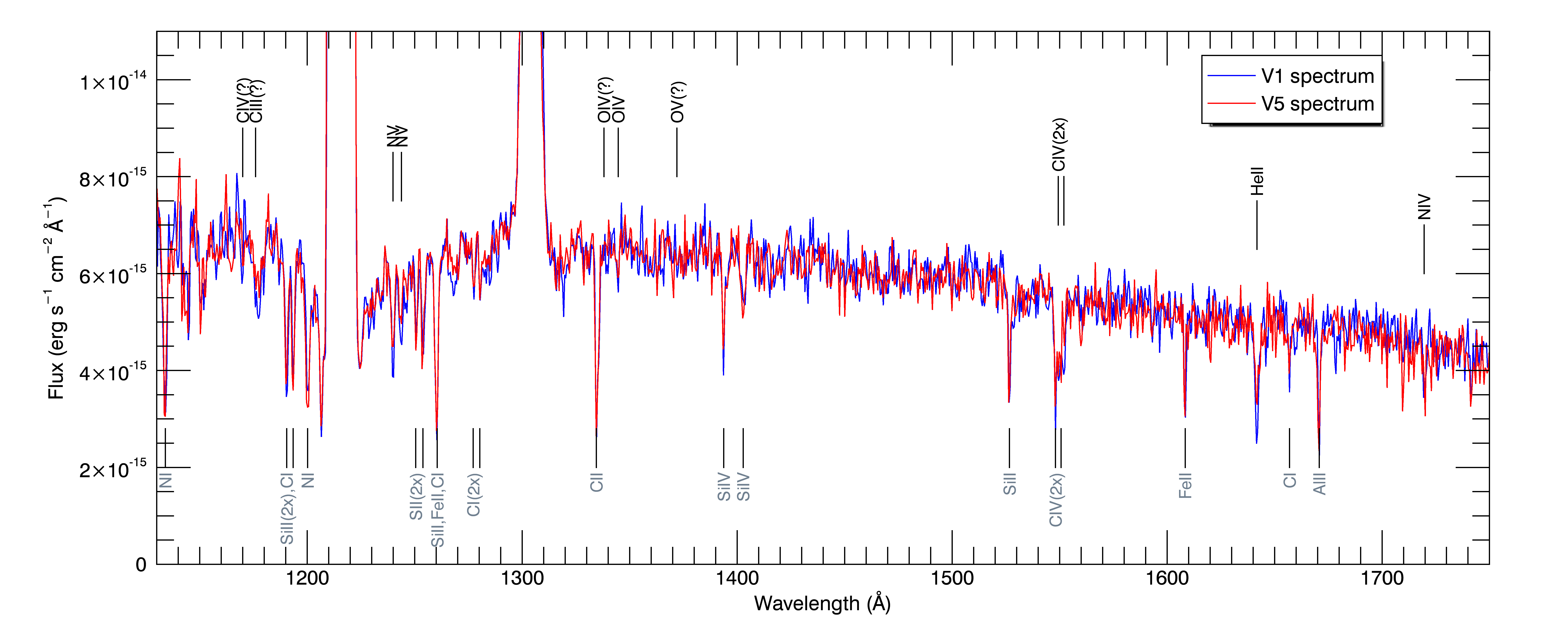}
\caption{COS spectra of V1 and V5 (binned over six pixels for more clarity), 
the flux from V5 was adjusted so that the two 
spectra overlap. The strongest interstellar lines are indicated below the 
spectra while the photospheric ones are labelled above. }
\label{specv1v5}
\end{center}
\end{figure*}

It is immediately apparent that 
several mass estimates are far too low to be reconciled with the
idea that most the hot subdwarfs are helium core or post-helium core
burning stars. While the six hottest stars in the sample, the hot H-rich
sdOs (including four pulsators), show a reasonable mean mass of 0.452 \msun, the 
rest of the
sample, taken as a whole, shows an unacceptably low mean mass value of
0.331 \msun, below the minimum mass for helium burning.
Interestingly, the spectroscopically inferred low mass problem for HB and EHB stars
in $\omega$ Cen has been encountered by \citet{moe11} and discussed in detail by
\citeauthor{moni2011} (\citeyear{moni2011,moni2012})\footnote{We note that 
\citeauthor{moni2011} and \citet{moe11} derived masses for their star in a 
different way: they used an empirical bolometric correction to determine the luminosity.}. 
The problem seems to affect only that cluster in particular. 

The mass distributions of the hot and cool subsamples are 
shown in Fig. \ref{mass}. The distributions were obtained following the procedure described in
\citet{fon12}: individual gaussians defined
by each individual value of the mass and its uncertainty are added together. 
Each gaussian has been normalised such that its surface
area is the same for each star, thus ensuring the same weight in the
addition procedure.
We do not believe the increase of mass with temperature to be 
real, especially considering that the hot sdO stars are likely to be post-EHB 
objects.
The mass difference would be naturally explained if the temperature 
of the sdO stars had been underestimated. Recomputing masses 
with temperatures increased by 10 000 K for the hottest stars leads to values compatible with the rest of the sample. 
We thus suggest that an underestimation of the effective 
temperature for the sdOs in the sample could explain the apparent mass 
discrepancy between the hot and cool samples.

\onltab{
\begin{table*}
\caption{Atmospheric and other parameters for the 38 stars in the sample of 
\citet{lat14}. The entries below the solid line refer to the ``hot'' subsample.}
\label{param}
\begin{tabular}{lccccc}
\hline
 Identifier & \teff\ (K) & log $g$ & log \nhe & $M_B$ & $M$/\msun \\
\hline
5238307 & 25711 $\pm$ 400  & 5.35 $\pm$ 0.06  & $-$2.27 $\pm$ 0.08 & 
4.428$\pm$0.147 & 0.231$\pm$0.102 \\
5139614 & 27594 $\pm$ 468  & 5.48 $\pm$ 0.06  & $-$3.74 $\pm$ 1.06 & 
4.661$\pm$0.147 & 0.211$\pm$0.102 \\
204071  & 28828 $\pm$ 602  & 5.53 $\pm$ 0.09  & $-$3.00 $\pm$ 0.14 & 
3.767$\pm$0.148 & 0.618$\pm$0.130 \\
168035  & 29770 $\pm$ 454  & 5.38 $\pm$ 0.07  & $-$3.27 $\pm$ 0.20 & 
3.767$\pm$0.147 & 0.401$\pm$0.111 \\
5262593 & 31161 $\pm$ 280  & 5.48 $\pm$ 0.05  & $-$3.07 $\pm$ 0.29 & 
3.648$\pm$0.146 & 0.531$\pm$0.092 \\
5243164 & 32403 $\pm$ 281  & 5.41 $\pm$ 0.05  & $-$2.65 $\pm$ 0.17 & 
4.044$\pm$0.146 & 0.242$\pm$0.093 \\
5180753 & 34850 $\pm$ 317  & 5.75 $\pm$ 0.06  & $-$1.46 $\pm$ 0.06 & 
4.238$\pm$0.147 & 0.466$\pm$0.101 \\
5142999 & 34477 $\pm$ 392  & 5.67 $\pm$ 0.07  & $-$1.09 $\pm$ 0.05 & 
4.387$\pm$0.146 & 0.308$\pm$0.110 \\
5222459 & 35008 $\pm$ 327  & 5.73 $\pm$ 0.05  & $-$0.73 $\pm$ 0.04 & 
4.663$\pm$0.146 & 0.253$\pm$0.093 \\
5119720 & 35018 $\pm$ 403  & 5.77 $\pm$ 0.07  & $-$0.81 $\pm$ 0.05 & 
4.457$\pm$0.147 & 0.381$\pm$0.110 \\
53945   & 35216 $\pm$ 316  & 5.91 $\pm$ 0.05  & $-$0.61 $\pm$ 0.04 & 
4.562$\pm$0.146 & 0.493$\pm$0.092 \\
75981   & 35929 $\pm$ 307  & 5.71 $\pm$ 0.05  & $-$1.05 $\pm$ 0.05 & 
4.151$\pm$0.146 & 0.432$\pm$0.093 \\
5164025 & 36020 $\pm$ 428  & 5.84 $\pm$ 0.07  & $-$0.55 $\pm$ 0.05 & 
4.716$\pm$0.146 & 0.315$\pm$0.110 \\
5205350 & 36251 $\pm$ 335  & 5.54 $\pm$ 0.06  & $-$0.61 $\pm$ 0.04 & 
4.300$\pm$0.146 & 0.193$\pm$0.101 \\
5165122 & 36331 $\pm$ 328  & 5.71 $\pm$ 0.06  & $-$0.64 $\pm$ 0.04 & 
4.439$\pm$0.146 & 0.288$\pm$0.101 \\
165943  & 36479 $\pm$ 401  & 5.76 $\pm$ 0.07  & $-$0.68 $\pm$ 0.05 & 
4.351$\pm$0.147 & 0.383$\pm$0.110 \\
5141232 & 36583 $\pm$ 402  & 5.72 $\pm$ 0.07  & $-$0.61 $\pm$ 0.05 & 
4.416$\pm$0.146 & 0.300$\pm$0.110 \\
274052  & 36640 $\pm$ 506  & 5.59 $\pm$ 0.09  & $-$0.35 $\pm$ 0.06 & 
4.558$\pm$0.147 & 0.140$\pm$0.130 \\
5242504 & 36653 $\pm$ 387  & 5.75 $\pm$ 0.07  & $-$0.45 $\pm$ 0.05 & 
4.495$\pm$0.146 & 0.298$\pm$0.110 \\
264057  & 36696 $\pm$ 408  & 5.70 $\pm$ 0.07  & $-$0.80 $\pm$ 0.05 & 
4.421$\pm$0.147 & 0.280$\pm$0.111 \\
5142638 & 36740 $\pm$ 428  & 5.71 $\pm$ 0.07  & $-$0.38 $\pm$ 0.05 & 
4.649$\pm$0.146 & 0.210$\pm$0.110 \\
5102280 & 36948 $\pm$ 327  & 5.70 $\pm$ 0.06  & $-$0.94 $\pm$ 0.05 & 
4.195$\pm$0.146 & 0.374$\pm$0.101 \\
177711  & 37093 $\pm$ 433  & 5.72 $\pm$ 0.07  & $-$0.45 $\pm$ 0.05 & 
4.429$\pm$0.147 & 0.287$\pm$0.111 \\
5220684 & 37544 $\pm$ 368  & 5.82 $\pm$ 0.07  & $-$0.86 $\pm$ 0.05 & 
4.338$\pm$0.147 & 0.435$\pm$0.110 \\
5062474 & 37554 $\pm$ 863  & 5.90 $\pm$ 0.14  & $-$0.09 $\pm$ 0.09 & 
4.757$\pm$0.148 & 0.330$\pm$0.183 \\
5138707 & 37855 $\pm$ 599  & 5.93 $\pm$ 0.09  &  0.57 $\pm$   0.05 & 
4.932$\pm$0.147 & 0.283$\pm$0.130 \\
5124244 & 38432 $\pm$ 530  & 5.97 $\pm$ 0.09  & $-$0.01 $\pm$ 0.05 & 
4.730$\pm$0.146 & 0.407$\pm$0.130 \\
5170422 & 38533 $\pm$ 340  & 5.60 $\pm$ 0.06  & $-$0.77 $\pm$ 0.04 & 
4.190$\pm$0.146 & 0.240$\pm$0.101 \\
5047695 & 38578 $\pm$ 549  & 5.69 $\pm$ 0.12  & $-$0.18 $\pm$ 0.07 & 
4.806$\pm$0.150 & 0.106$\pm$0.163 \\
5085696 & 39072 $\pm$ 371  & 5.66 $\pm$ 0.08  & $-$0.04 $\pm$ 0.05 & 
4.585$\pm$0.146 & 0.147$\pm$0.120 \\
5039935 & 39804 $\pm$ 523  & 6.06 $\pm$ 0.11  &  0.49 $\pm$ 0.07   & 
4.998$\pm$0.148 & 0.353$\pm$0.151 \\
165237  & 43843 $\pm$ 362  & 6.01 $\pm$ 0.11  &  0.75 $\pm$ 0.10   & 
4.173$\pm$0.147 & 0.579$\pm$0.149 \\
\hline
5242616 & 44959 $\pm$ 637  & 5.88 $\pm$ 0.08  & $-$1.41 $\pm$ 0.08 & 
4.265$\pm$0.146 & 0.392$\pm$0.120 \\
5034421 (V1) & 49113 $\pm$ 824  & 5.89 $\pm$ 0.07  & $-$1.76 $\pm$ 0.09 & 
4.137$\pm$0.147 & 0.408$\pm$0.112 \\
177238 (V3) & 49328 $\pm$ 877  & 6.07 $\pm$ 0.08  & $-$1.73 $\pm$ 0.11 & 
4.138$\pm$0.147 & 0.621$\pm$0.121 \\
154681 (V4) & 50635 $\pm$ 758  & 5.89 $\pm$ 0.08  & $-$1.25 $\pm$ 0.05 & 
4.043$\pm$0.146 & 0.434$\pm$0.120 \\
281063 (V5) & 58789 $\pm$ 1910  & 6.12 $\pm$ 0.11  & $-$1.67 $\pm$ 0.13 & 
4.344$\pm$0.147 & 0.487$\pm$0.152 \\
177614  & 59724 $\pm$ 1288  & 6.02 $\pm$ 0.08  & $-$1.32 $\pm$ 0.09 & 
4.281$\pm$0.147 & 0.392$\pm$0.120 \\
\hline
\end{tabular}
\end{table*}
}

\section{The UV analysis}

\subsection{Observations}
Ultraviolet COS spectra of the two pulsators V1 and V5 were obtained during 
cycle 22 (proposal GO-13707). Each star was observed for 5337 s with the low 
resolution G140L (R $\approx$3000) grating in time-tag mode. The data were reduced 
following the standard CALCOS procedure.

The resulting spectrograms of both stars are shown in Fig. \ref{specv1v5}, where 
the flux of V5 was multiplied by a factor 1.138 in order to match the flux of 
V1, thus emphasing the similarity between the spectra of both stars. This is 
somewhat expected given that the pulsators form a rather homogeneous group 
according to the optical analysis. We can also note that the shape of the 
continuum is essentially the same for both stars. Since this is largely due to interstellar 
reddening, it is not surprising that it is very similar for the two targets. 
Most of the strong spectral features are due to the interstellar medium and are 
labelled below the spectra in Fig.  \ref{specv1v5}; C~\textsc{i-ii}, \ion{N}{i} 
\ion{Si}{II}, \ion{Fe}{II} and \ion{Al}{II}, also conspicuous are the Ly$\alpha$ 
and \ion{O}{i} ($\lambda$1304) geocoronal emission lines. The \ion{Si}{iv} 
resonance doublet ($\lambda\lambda$1394,1403) is also visible in both stars at a 
radial velocity (RV) consistent with zero, thus indicating its interstellar 
origin. No stellar component can be resolved for this doublet. As for the 
\ion{C}{iv} resonance lines ($\lambda\lambda$1548,1551), both the interstellar 
(RV $\approx\-$20 km s$^{-1}$) and photospheric components (RV $\approx$230 km 
s$^{-1}$) can be resolved. Additional photospheric lines are the \ion{N}{v} 
resonance doublet ($\lambda\lambda$1239,1243), \ion{N}{iv} $\lambda$1718 and 
\ion{He}{ii} $\lambda$1640, which are found at a radial velocity consistent with 
that of the cluster (232 km s$^{-1}$, \citealt{har96}).

\subsection{The COS light curves}

\begin{figure*}[t]
\begin{center}
\includegraphics[scale=0.45]{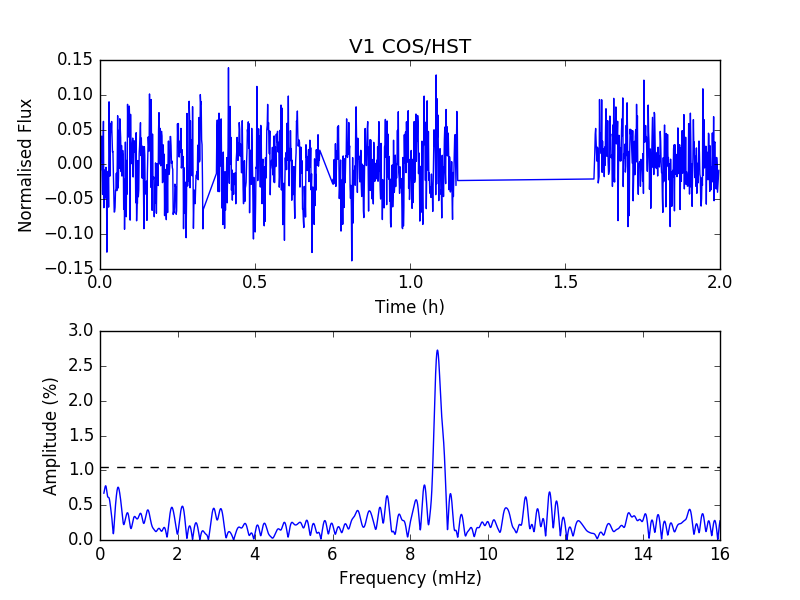}
\includegraphics[scale=0.45]{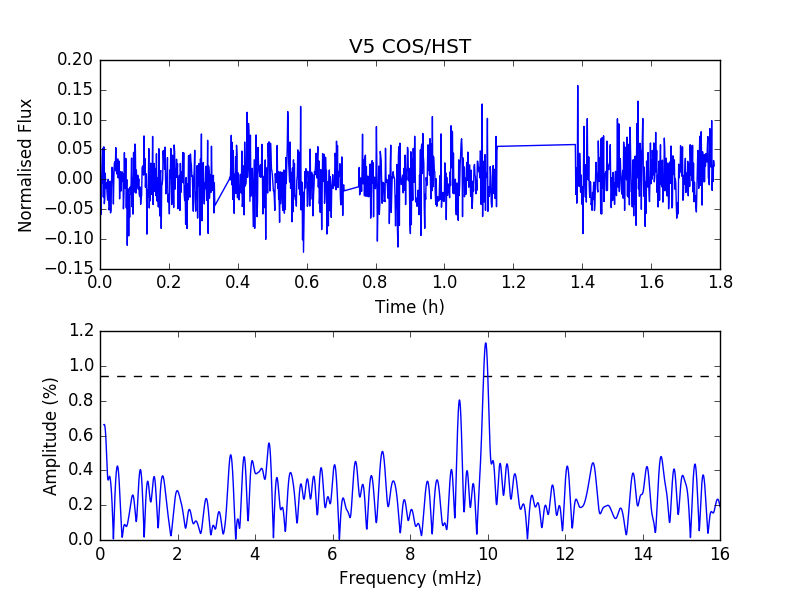}
\caption{Light curves and Fourier amplitude spectrum corresponding to the COS 
observations of V1 and V5. The horizontal dashed line indicates the 3.5$\sigma$ 
detection threshold.}
\label{lc}
\end{center}
\end{figure*}

We took advantage of the time-tag mode to construct the UV light curves of both 
stars using the LightCurve tool\footnote{http://justincely.github.io/lightcurve/}, a discussion
on the method is presented in \citet{sand16}.
The time-tag counts were binned into datachunks of 5 s. This is small enough to fully resolve the expected pulsations in the $\approx$80-120 s range while still giving adequate S/N in each data point.

The top panels of Figure \ref{lc} show the light curves obtained for the two stars, normalised to the average
flux of the star in question. They are each divided into four chunks of continuous data, corresponding to observations 
at the four COS FPOS positions. The large gap in each curve represents the re-acquisition period.   

The lower panels of Figure \ref{lc} show the Fourier amplitude spectrum based on the COS lightcurves. 
Periodicities with amplitudes above the 3.5 $\sigma$ threshold were extracted at 114.8 s (2.7\% amplitude) 
and 113.4 s (1.2\% amplitude) for V1, and at 100.5 s (1.1\% amplitude) for V5. These periods correspond 
 well to the dominant modes known for these stars from ground-based optical time-series photometry
\citep{ran16,ran11}. For V5, an additional period known from the optical data is also recovered just 
below 3.5 $\sigma$ at 107.8 s (0.6\% amplitude).

While it is interesting to compare the observed periodicities and amplitudes derived from the optical and 
the COS data at the qualitative level, a quantitative comparison is not particularly instructive due to the 
very different time baselines of the datasets. Indeed, for V1 the optical $u'$-band light curve obtained over a 
6-day period in 2009 yields a rather messy Fourier spectrum with several closely split components clustered around 
115.0 s, 114.7 s and 114.4 s \citep[see Table 4,][]{ran16}, whereas an earlier $B$ band dataset taken over two nights 
in 2009 shows dominant periods at 114.7 s and 113.7 s \citep[see Table 1,][]{ran11}, very similar to those 
uncovered with COS. It is not clear which of the split components constitute independent harmonic oscillations and
which are induced in the Fourier spectrum, for example by intrinsic amplitude variations or the beating of closely spaced modes.
The longer the time baseline and the better the quality of the data, the more complicated the Fourier spectrum 
appears to become. For V5 the situation seems simpler, the previously extracted 100.6 s, 99.3 s and 107.5 s 
periodicities \citep{ran16} not being split in any obvious way, but this is likely due to the much lower
 pulsational amplitudes and the limited S/N of the data. 

Given the clear indications for pulsational amplitude variability there is no point in trying to use the 
relative amplitudes observed in the different frequency bands for mode identification. If simultaneous
time-series photometry were available it would in principle be possible to exploit the colour dependence of 
each mode's amplitude to derive the degree index $\ell$, as has been done with some success for rapidly pulsating 
sdB stars in the field \citep{ran05}. Since in our case the different datasets are taken several years apart 
 we limit ourselves to a very qualitative comparison of the apparent amplitudes in the different bands. For 
the low-degree modes expected to be observed in these stars we would expect a general trend of amplitude 
decrease with increasing wavelength, that is, the apparent COS far-UV amplitudes should be significantly higher than those 
seen in the optical data (assuming a similar intrinsic amplitude of the mode at the time of observation) due to 
the frequency-dependence of the limb darkening \citep[see][for details]{ran05}. This is assuming these stars 
behave similarly to the much cooler pulsating sdB stars, since detailed calculations of the pulsational perturbation 
of the stellar atmosphere have not yet been carried out for sdO stars. 

In our very rudimentary colour-amplitude analysis we simply add up the amplitudes of all frequency components around the 115 s complex for V1. This gives a far-UV amplitude of 3.9\%, a $u'$ amplitude of 5.1\% and a $B$ amplitude of 2.2\%. For the 100.6 s pulsation in V5 we find 1.1\% in the far-UV vs 0.54 \% in the $u'$, and for the 107.5 s pulsation we have 0.6\% in the far-UV and 0.42 \% in the $u'$ (no $B$-band data are available for this star).
With the exception of the very large $u'$ amplitude derived for the 115 s complex in V1 (that value is particularly unreliable due to the many split components), the general trend does seem to be for the UV amplitudes to be larger than the corresponding peaks in the optical. However, given the small number statistics this is not a significant result, and it is not clear how indicative the amplitudes derived from the $u'$ and $B$ bands are. We would expect the amplitude from the ground-based optical data to underestimate the true apparent
amplitudes due to the flux contribution from nearby (presumably not pulsating) stars in the very crowded $\omega$ Cen field. This is true in particular for V5, which has a very close relatively bright companion. We tentatively conclude that the far-UV amplitudes observed in the $\omega$ Cen variables appear qualitatively comparable to or up to twice as large as those from ground-based optical data. This finding is of interest in the context of comparing the far-UV space-based pulsational properties of hot subdwarfs (such as those obtained for the pulsators in NGC2808 by \citealt{bro13}) to those derived from ground-based optical data.

\begin{figure*}
\begin{center}
\includegraphics[scale=0.46]{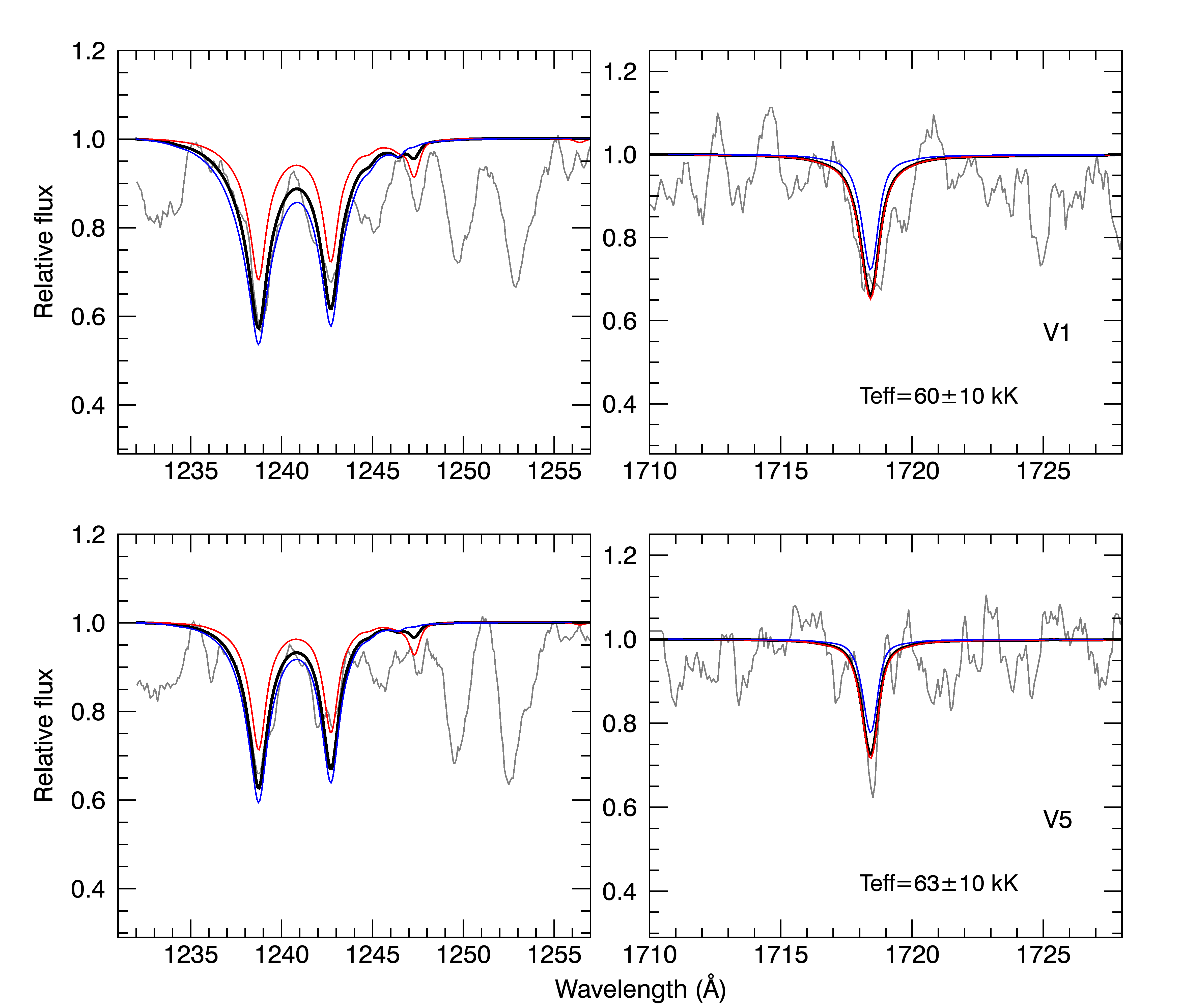}
\includegraphics[scale=0.46]{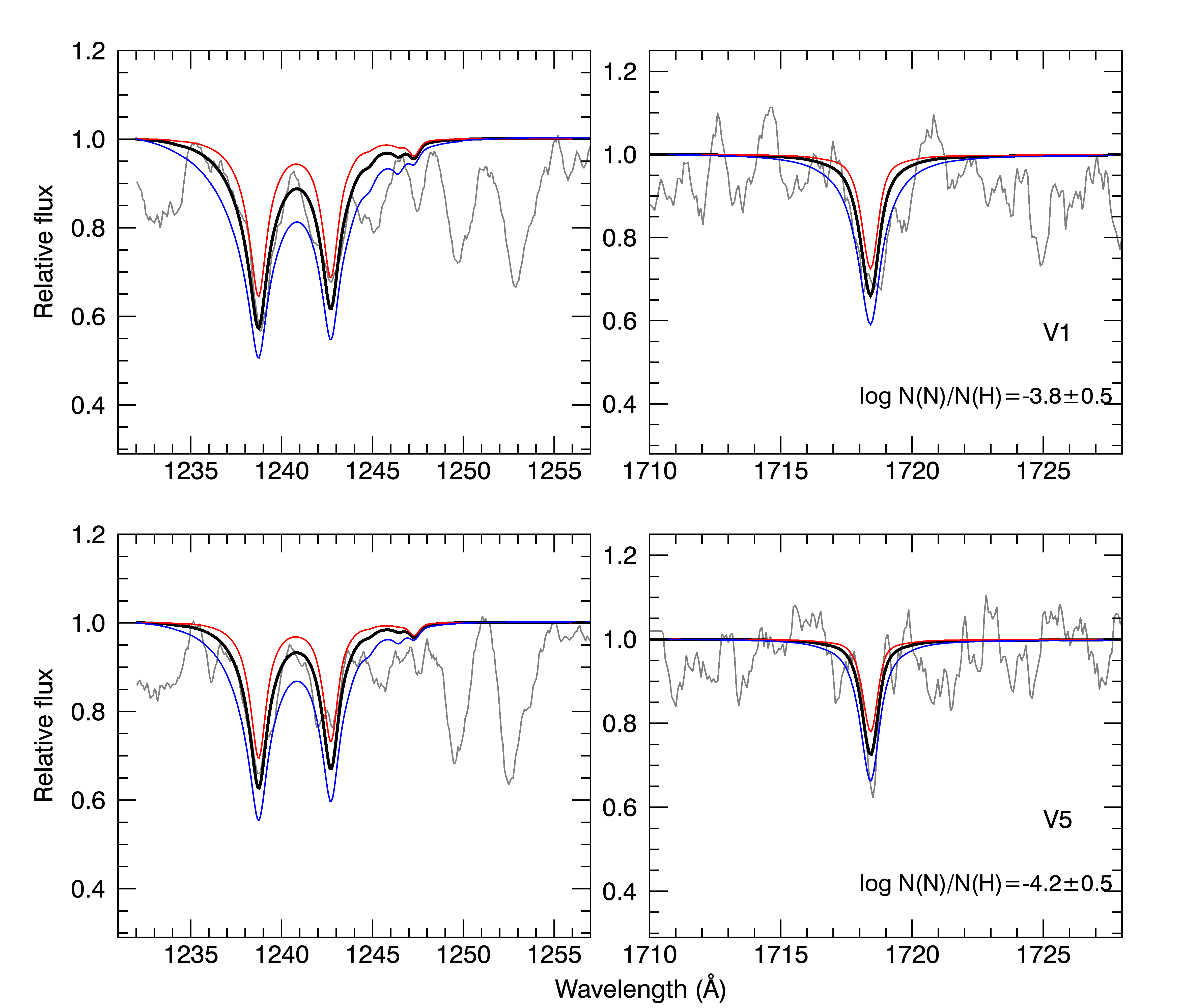}
\caption{Best fitting models (thick black line) for the \ion{N}{v} doublet 
and \ion{N}{iv} line in the spectrum of V1 (upper panels) and V5 
(lower panels). The observed spectra 
(grey) are smoothed over a six pixels width (one resolution element) in this
figure and the following ones.
Right panels:  As a comparison, model spectra with a lower (red) and higher (blue) temperature are shown.
Left panels: With model spectra having a lower (red) and higher (blue) nitrogen abundance.  }
\label{fitn}
\end{center}
\end{figure*}

\subsection{Analysis of the COS spectra}

Our goal was to use strong metal lines in the UV range to simultaneously 
determine the metal abundance and the temperature, using lines originating from 
different ionization levels. In the observed wavelength range ($\approx$1150$-$2000 
\AA), lines originating from C~\textsc{iii}-\textsc{iv}, 
N~\textsc{iv}-\textsc{v} and O~\textsc{iv}-\textsc{v} are predicted. As already 
mentioned, we detect the nitrogen lines and the \ion{C}{iv} doublet in our 
spectra. Unfortunately, the \ion{C}{iii} multiplet ($\approx$1176 \AA) and the 
\ion{C}{iv} lines (1169 \AA) cannot be used for the analysis. 
A wide 
absorption feature between $\approx$1170-1180 \AA\ is seen in both spectra (and 
indicated in Fig. \ref{specv1v5}), hiding any hints of the \ion{C}{iii} 
multiplet. We could not find the origin of this feature. No strong interstellar 
lines are expected in this region and we 
verified that no artefact was produced when combining the four subexposures. 
Iron and nickel lines can be abundant at these wavelengths but are not expected 
to produce such a strong feature. To nevertheless examine this possibility we compared an 
IUE spectrum (degraded to the COS resolution) of Feige 34 with our COS spectra, but
did not find any similar feature. Feige 34 is an sdO star with parameters, to 
our best knowledge, similar to those of V5 but enriched in iron and nickel \citep{lat16}. 
If those lines were at the origin of the feature seen in the COS spectra, we 
would expect it to feature even more prominently in the IUE spectrum of Feige 34. 
Although the COS handbook mentions calibration issues at these wavelengths, the same phenomenon was seen by \citet{bro12} (see their figure 12) with STIS/G140L spectra of EHB stars, suggesting that there might be an unknown astrophysical explanation.
Finally, a firm identification of the oxygen lines (1338 \AA, 1342 \AA, and 1371 \AA) 
was not possible, suggesting a sub-solar abundance. Thus we had 
to rely on the nitrogen lines for our analysis. 

For each star we built a two-dimensional grid of NLTE model atmospheres, varying 
\teff\ and the nitrogen abundance, while keeping the other parameters fixed. 
For both stars the grid covered the following ranges: \teff\ from 45 000 K to 65 
000 K in steps of 2 000 K and log $N$(N)/$N$(H) from $-$3.0 to $-$5.8 in 0.4 dex intervals. 
The surface gravity and helium abundance were fixed for each star to the values 
derived by the optical analysis (see Table \ref{param}). The model atmospheres 
and synthetic spectra were computed with the public codes TLUSTY and SYNSPEC 
\citep{lanz03,lanz07}, and include C, N, and O as metallic elements in addition 
to H and He. 
We adopted in these models a carbon abundance of log $N$(C)/$N$(H)=$-$4.6, according
to the estimates from the optical spectra \citep{lat14}, and an oxygen abundance of one tenth solar, given the weakness of the observed oxygen features.
Because the line spread function of COS departs from the usual 
Gaussian function, we used the theoretical profiles listed on the COS website 
\footnote{http://www.stsci.edu/hst/cos/performance/spectral\_resolution} 
(G140L/1105 at lifetime position 3) for the convolution of our synthetic spectra. 

We then used our synthetic spectra grids to simultaneously fit the nitrogen 
lines in the observed spectra. The result can be seen in Fig. \ref{fitn}, where 
the thin grey line represents the observation, while the thick black line is a 
best fitting model. In the case of V1 we obtained values of log $N$(N)/$N$(H) = 
$-$3.8$\pm$0.5 dex and \teff\ = 60 000 $\pm$ 5 000 K. The nitrogen lines in V5 
being less strong, the fit for this star resulted in a lower N abundance of log 
$N$(N)/$N$(H) = $-$4.2$\pm$0.5 dex combined with \teff\ = 63 000 $\pm$ 6 000 
K.
The line profiles are not very well defined at low resolution, and the SNR is 
relatively low in the 1718 \AA\ region, which explains the rather high uncertainties. 
Nevertheless, the nitrogen lines indicate effective temperatures higher 
than the optical estimates (by $\approx$10 000 K for V1 and $\approx$5 000 K for V5). 
The nitrogen abundances are consistent with a solar value ($-$4.2) which 
is rather typical for hot subdwarf stars \citep{bla08,geier13}.
For comparison, in Fig. \ref{fitn} we overplot the N line 
profiles for temperatures of $\pm$10 kK than those obtained with the fit on the left panels.
As seen from the 
figure, the \ion{N}{iv} line is not sensitive to temperature below $\approx$63~kK, while the 
\ion{N}{v} doublet appears too weak at lower \teff. At the lower temperatures, 
the higher N abundance required to match the \ion{N}{v} lines would produce a 
\ion{N}{iv} line stronger than observed. The effect of changing the nitrogen abundance by
$\pm$0.5 dex is shown on the right panels of Fig. \ref{fitn}.

\begin{figure}
\resizebox{\hsize}{!}{\includegraphics{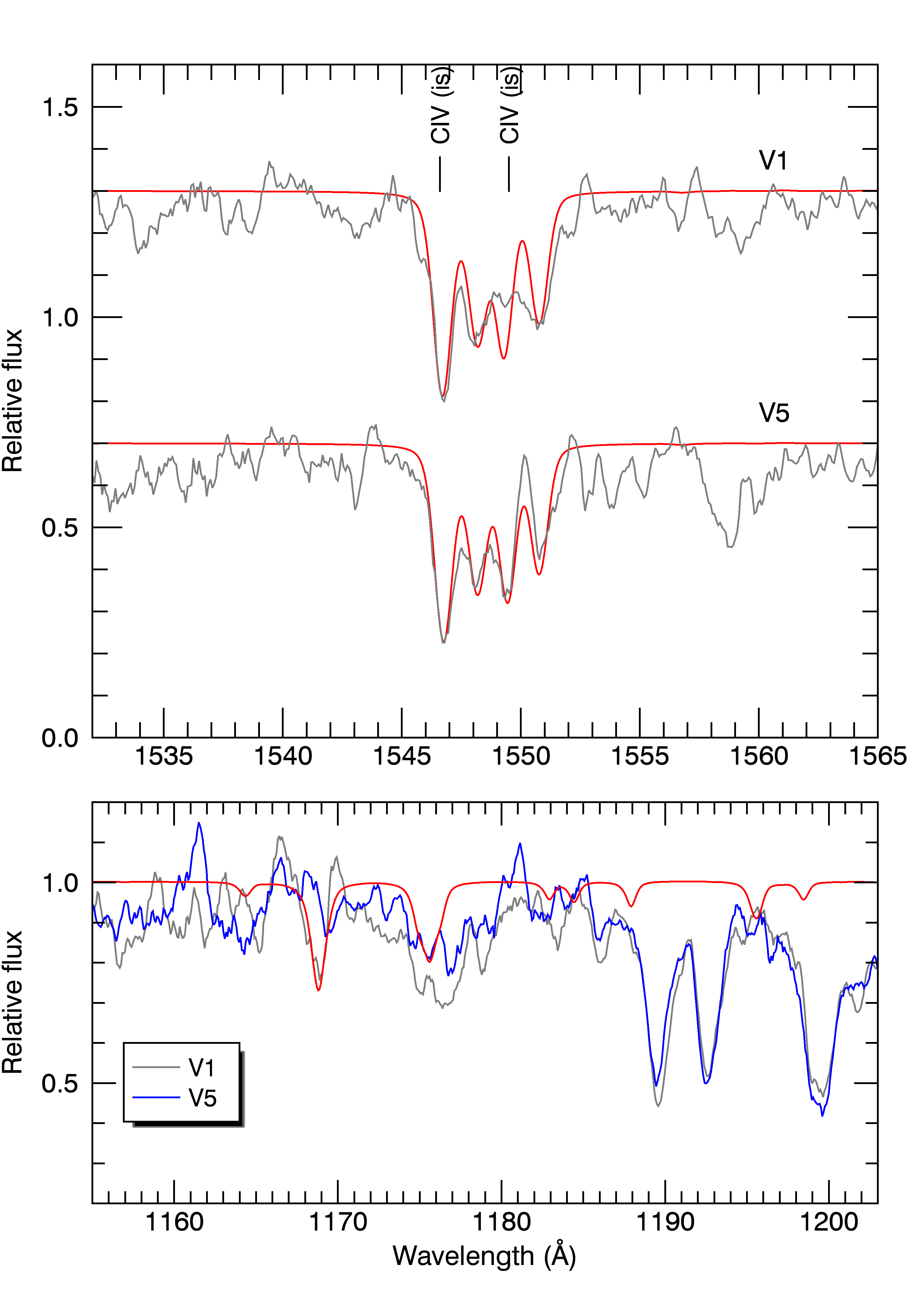}}
\caption{Upper panel : 
Comparison between the observed (grey) spectra of V1 and V5 in the 
\ion{C}{iv} doublet region and synthetic spectra combining interstellar 
and photospheric lines (red). The normalized spectra 
are shifted along the $y$ axis for better visualization.
Lower panel:
The region of the \ion{C}{iv} (1169 \AA) and \ion{C}{iii} ($\approx$1176 \AA) lines
is shown for V1 and V5. A theoretical spectrum is shown in red as comparison.
}
\label{V1V5_C}
\end{figure}

Similar temperature $-$ abundance grids were made for oxygen and carbon in order 
to estimate their abundances.
We examined the 1330-1380 \AA\ region where the strongest oxygen lines are 
predicted. The only line that could be identified in both stars 
is the \ion{O}{iv} at 1342 \AA. The other lines are not adequately defined to
claim a real detection. Nevertheless, we can set an upper limit on log $N$(O)/$N$(H)
$\approx$ $-$4.6. We recall 
here the solar abundance to be $-$3.3 dex, so the stars would have abundances 
below 1/10 solar, which is also a normal value for hot subdwarfs.
Considering carbon, although the IS and photospheric components of the \ion{C}{iv} 
doublet can be distinguished, the line profiles are blended. However, the line 
strength remains in agreement with the upper limits placed from the optical 
spectra, log $N$(C)/$N$(H)$\lesssim$ $-$4.6 (no carbon lines are 
distinguishable in the optical). The upper panel of Fig. \ref{V1V5_C} shows the observed and 
modelled carbon doublet (interstellar and stellar components)
for V1 and V5. In the lower panel, we show the observed spectra in the $\approx$1175 \AA\ region as well as 
a theoretical spectrum representative for both stars.
The issue of the \ion{C}{iii} multiplet has been discussed previously, and the plot clearly illustrates that the 
predicted carbon line cannot explain the observed feature around 1175 \AA.

\begin{figure}
\resizebox{\hsize}{!}{\includegraphics{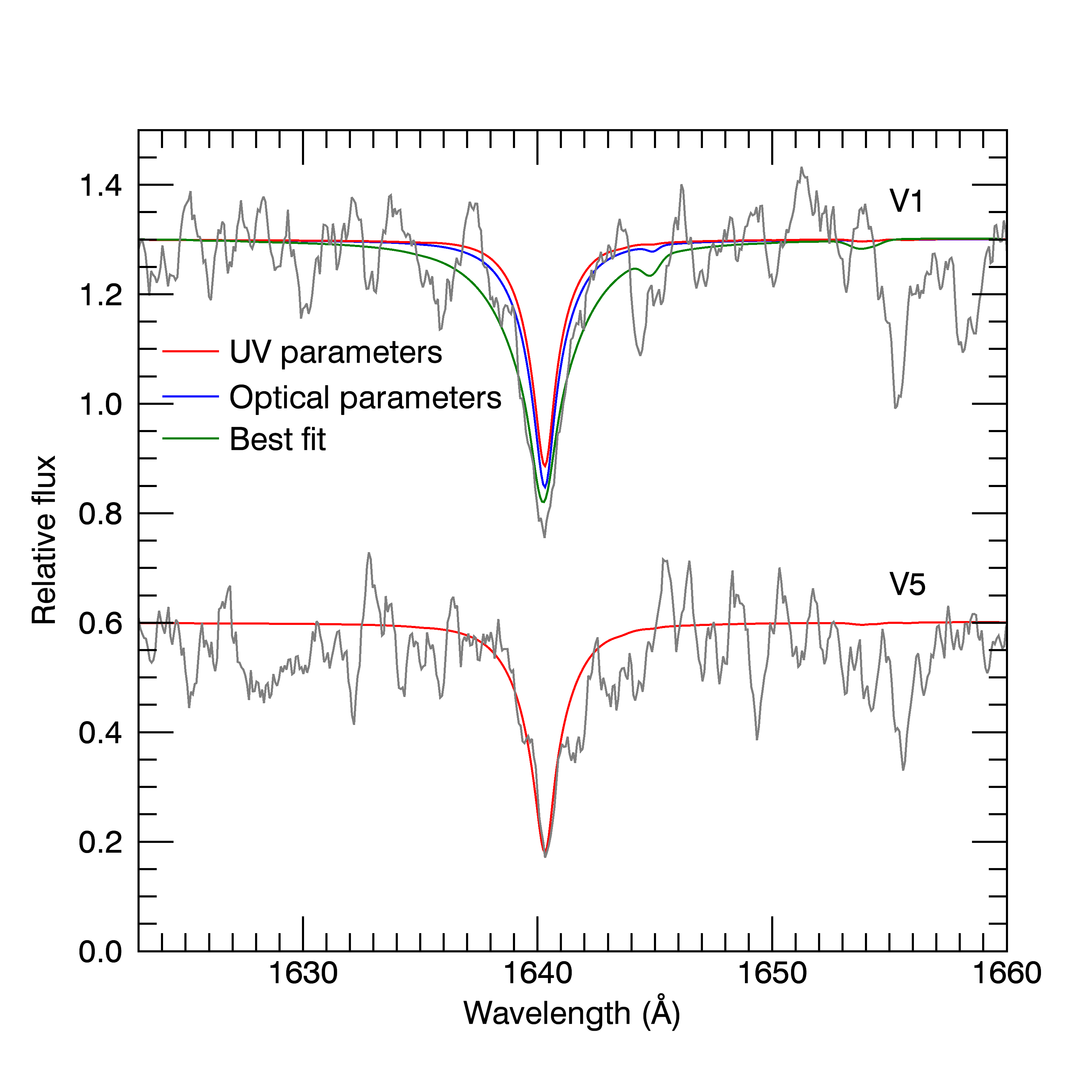}}
\caption{Comparison between the observed \ion{He}{ii} $\lambda$1640 line (grey) 
for V1 and V5 and synthetic spectra at their UV optimal temperature (red). For 
V1 we added also a model at the optical determined temperature (blue) and the 
best fit model to the line (green). }
\label{heII}
\end{figure}

\begin{figure}[t]
\resizebox{\hsize}{!}{\includegraphics{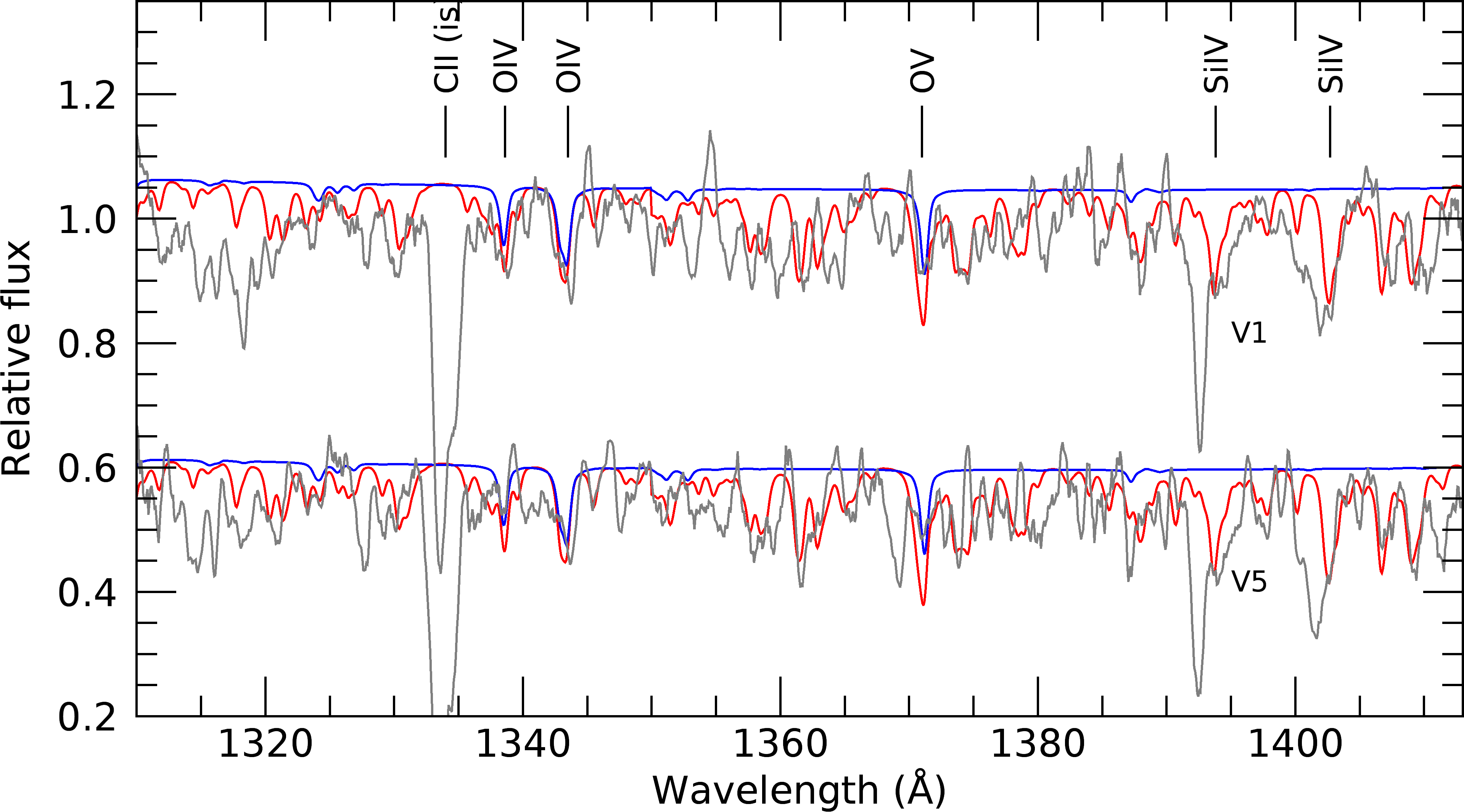}}
\caption{COS spectra of V1 and V5 (grey) compared with model spectra including C, N, and O at their derived values 
(blue) and an additional solar amount of Fe and Si (red)}.
\label{V5_1330}
\end{figure}

The final photospheric feature visible in the spectra is the \ion{He}{ii} line at 
1640 \AA. Figure \ref{heII} shows the region surrounding this line in the COS 
spectra of V1 and V5, overplotted (red) by synthetic models with a \teff\ of 
60 and 63 kK respectively. 
For V5, we have a good match between the synthetic and observed spectra, while 
for V1 the observed \ion{He}{ii} line is too strong to be reproduced by the 
synthetic spectrum.
This is somewhat unexpected since the helium abundance derived from the optical 
spectra is very similar for both stars (log \nhe\ = $-$1.76 and $-$1.67 for 
V1 and V5 respectively). The effective temperature affects the strength of the He 
lines, but in the \teff\ range of our stars the effect is not very strong for 
$\lambda$1640. This is illustrated with the blue curve in Fig. \ref{heII} which 
shows the He line at \teff\ = 49 000 K (the value derived from the optical 
spectra, see Table \ref{param}). The \ion{He}{ii} line is indeed stronger at 
lower \teff, but still not strong enough to match the observed profile which is 
wider and deeper still. 
We also fit the \ion{He}{ii} line of V1 in the temperature$-$helium 
abundance parameter space. The resulting best fit is illustrated by the green 
curve in the plot, corresponding to the following parameters: \teff\ = 51 000 
$\pm$ 4000 K and log \nhe\ = $-$0.76 $\pm$ 0.3. This helium abundance is ten 
times higher than the one indicated by the optical spectrum and thus in 
serious contradiction; such an abundance would produce optical helium lines that are much
stronger than observed.
Moreover the theoretical line profile does not fit the UV 
spectrum very well, the wings being too broad and the core not deep enough. 
To conclude, we do not understand the \ion{He}{ii} line in the COS spectrum of V1. 

As mentioned previously, the silicon resonance lines have RVs indicating an 
interstellar origin with no resolved photospheric components. Individual lines 
from heavy elements like iron and nickel can also not be resolved at the low 
resolution and relatively poor S/N of our spectra. However, by comparing the COS 
spectra with model spectra including a solar amount of silicon and iron, 
we found such abundances to be compatible with our observations. Thus, a
solar abundance is a reasonable upper limit to place on these elements.

Figure \ref{V5_1330} displays our observed spectra over a spectral range especially 
rich in \ion{Fe}{v} lines. Overplotted in red are model spectra for both stars including silicon 
and iron at solar abundances (C, N, O are also included at their estimated 
values). Oxygen and silicon lines are indicated, while the other features are 
due to iron, as comparison, model spectra without Si and Fe are plotted in blue. 
We note that the \ion{O}{v} $\lambda$1370 line is blended with iron lines present at its blue side, 
which makes it appear stronger in a model including iron.

\section{Conclusion}

We addressed the issue of the temperature discrepancy of the \omcen\ instability 
strip discussed in \citet{ran16}. As shown in their Figure 11, the pulsating 
sdOs are found at temperatures cooler than the predicted instability region 
favourable to pulsations. The discrepancy could be explained either by 
shortcomings in the seismic models (e.g. nickel opacity might boost pulsations 
at lower effective temperatures) or by an underestimation of the surface 
temperature of the stars derived by means of optical spectroscopy. 
In this paper, we focussed on the second possibility by analysing low
resolution COS spectra of two pulsators (V1 and V5). We also determined a mass 
distribution for the sample of EHB stars presented in \citet{lat14}.

From the mass distribution, we noticed a significant difference between the mean 
mass of the "cool" stars of the sample (with \teff $\lta$ 45 kK) and the hotter sdOs.
The latter group includes only six stars (including four of the 
pulsators) but the mean mass difference of 0.127 \msun\ is nevertheless 
suspicious. This mean mass discrepancy could indicate that the effective 
temperatures are underestimated from the optical spectra, since an increase 
in the temperature of the stars 
would lead to a lower radius (and mass) needed to match the observed magnitude 
of the stars, given a fixed distance to the cluster.

To investigate the issue in more detail, we analysed low resolution 
UV spectra for two of the pulsators, namely V1 and V5. The goal was to use the 
ionization equilibrium of strong metal lines (C, N, O) originating from 
different ionic states to assess more precisely the effective temperature of the 
stars. However, only the nitrogen lines could be used to this purpose.
From the \ion{N}{v} doublet and \ion{N}{iv} $\lambda$1718 lines we could estimate
the nitrogen abundances to be close to solar for V5 and slightly higher in V1.
Given these nitrogen abundances, 
the observed strength of the \ion{N}{v} doublet indicates temperatures significantly higher 
than those estimated from optical spectroscopy. For V1 we obtained \teff\ = 60$\pm$5 kK and a slightly 
higher temperature of \teff\ = 63$\pm$6 kK for V5. 
The uncertainties are rather large given 
that we are forced to rely on three spectral lines in low quality spectra, but 
it is nevertheless quite clear that the nitrogen lines require the effective 
temperature to be closer to 60 kK than 50 kK for both stars. As for other 
elements, the strength of the \ion{C}{iv} doublet is consistent with the upper 
limit derived from the optical spectra of about 1/10 solar and the oxygen 
abundance is also depleted by at least a factor of 10 with respect to solar.
Upper limits for silicon and iron are about solar.

In summary, based on the mass distribution of the \omcen\ sdOs, and the UV 
nitrogen lines of the two pulsating stars V1 and V5, it is likely that the 
effective temperatures of the sdOs in the $\omega$ Cen sample are systematically higher than those 
derived from the optical spectra, thus moving the stars into the instability region predicted by 
seismic models. Our analysis was however limited by the quality of the data available.
The analysis and parameter determination of hot stars (sdOs, white dwarfs) is hampered by
 inherent difficulties, such as the behaviour of the Balmer 
lines and the weak temperature dependence of the UV and optical flux distribution. 
Combining these issues with the observational difficulties faced for faint stars in a 
very crowded field makes studies such as that attempted here very challenging indeed.
We believe the most promising way forward is to conduct spectroscopic studies of 
brighter stars with effective temperatures similar to 
the \omcen\ sdOs. This could go a long way towards an understanding of the fundamental parameters of these stars 
and how to reliably derive them, as well as their chemical patterns and 
evolutionary status.

\begin{acknowledgements}
This work was supported by a fellowship for postdoctoral
researchers from the Alexander von Humboldt Foundation awarded to M.L., who also 
acknowledges funding by the Deutsches Zentrum für
Luft- und Raumfahrt (grant 50 OR 1315).
This research makes use of the SAO/NASA Astrophysics Data System Bibliographic Service.

\end{acknowledgements}

%
%

\bibliographystyle{aa}
\bibliography{refOmCen}

\begin{thebibliography}{43}
\expandafter\ifx\csname natexlab\endcsname\relax\def\natexlab#1{#1}\fi

\bibitem[{{Bill{\`e}res} {et~al.}(2002){Bill{\`e}res}, {Fontaine}, {Brassard},
  \& {Liebert}}]{bil02}
{Bill{\`e}res}, M., {Fontaine}, G., {Brassard}, P., \& {Liebert}, J. 2002,
  \apj, 578, 515

\bibitem[{{Blanchette} {et~al.}(2008){Blanchette}, {Chayer}, {Wesemael},
  {Fontaine}, {Fontaine}, {Dupuis}, {Kruk}, \& {Green}}]{bla08}
{Blanchette}, J.-P., {Chayer}, P., {Wesemael}, F., {et~al.} 2008, \apj, 678,
  1329

\bibitem[{{Braga} {et~al.}(2016){Braga}, {Stetson}, {Bono}, {Dall'Ora},
  {Ferraro}, {Fiorentino}, {Freyhammer}, {Iannicola}, {Marengo}, {Neeley},
  {Valenti}, {Buonanno}, {Calamida}, {Castellani}, {da Silva},
  {Degl'Innocenti}, {Di Cecco}, {Fabrizio}, {Freedman}, {Giuffrida}, {Lub},
  {Madore}, {Marconi}, {Marinoni}, {Matsunaga}, {Monelli}, {Persson},
  {Piersimoni}, {Pietrinferni}, {Prada-Moroni}, {Pulone}, {Stellingwerf},
  {Tognelli}, \& {Walker}}]{bra16}
{Braga}, V.~F., {Stetson}, P.~B., {Bono}, G., {et~al.} 2016, \aj, 152, 170

\bibitem[{{Brown} {et~al.}(2013){Brown}, {Landsman}, {Randall}, {Sweigart}, \&
  {Lanz}}]{bro13}
{Brown}, T.~M., {Landsman}, W.~B., {Randall}, S.~K., {Sweigart}, A.~V., \&
  {Lanz}, T. 2013, \apjl, 777, L22

\bibitem[{{Brown} {et~al.}(2012){Brown}, {Lanz}, {Sweigart}, {Cracraft},
  {Hubeny}, \& {Landsman}}]{bro12}
{Brown}, T.~M., {Lanz}, T., {Sweigart}, A.~V., {et~al.} 2012, \apj, 748, 85

\bibitem[{{Calamida} {et~al.}(2005){Calamida}, {Stetson}, {Bono}, {Freyhammer},
  {Grundahl}, {Hilker}, {Andersen}, {Buonanno}, {Cassisi}, {Corsi}, {Dall'Ora},
  {Del Principe}, {Ferraro}, {Monelli}, {Munteanu}, {Nonino}, {Piersimoni},
  {Pietrinferni}, {Pulone}, \& {Richtler}}]{cala05}
{Calamida}, A., {Stetson}, P.~B., {Bono}, G., {et~al.} 2005, \apjl, 634, L69

\bibitem[{{Charpinet} {et~al.}(1997){Charpinet}, {Fontaine}, {Brassard},
  {Chayer}, {Rogers}, {Iglesias}, \& {Dorman}}]{char97}
{Charpinet}, S., {Fontaine}, G., {Brassard}, P., {et~al.} 1997, \apjl, 483,
  L123

\bibitem[{{Del Principe} {et~al.}(2006){Del Principe}, {Piersimoni}, {Storm},
  {Caputo}, {Bono}, {Stetson}, {Castellani}, {Buonanno}, {Calamida}, {Corsi},
  {Dall'Ora}, {Ferraro}, {Freyhammer}, {Iannicola}, {Monelli}, {Nonino},
  {Pulone}, \& {Ripepi}}]{delp06}
{Del Principe}, M., {Piersimoni}, A.~M., {Storm}, J., {et~al.} 2006, \apj, 652,
  362

\bibitem[{{Dixon} {et~al.}(2016){Dixon}, {Chayer}, \& {Benjamin}}]{dix16}
{Dixon}, W.~V.~D., {Chayer}, P., \& {Benjamin}, R.~A. 2016, in American
  Astronomical Society Meeting Abstracts, Vol. 227, American Astronomical
  Society Meeting Abstracts, 239.05

\bibitem[{{Fontaine} {et~al.}(2012){Fontaine}, {Brassard}, {Charpinet},
  {Green}, {Randall}, \& {Van Grootel}}]{fon12}
{Fontaine}, G., {Brassard}, P., {Charpinet}, S., {et~al.} 2012, \aap, 539, A12

\bibitem[{{Fontaine} {et~al.}(2008{\natexlab{a}}){Fontaine}, {Brassard},
  {Green}, {Chayer}, {Charpinet}, {Andersen}, \& {Portouw}}]{font08}
{Fontaine}, G., {Brassard}, P., {Green}, E.~M., {et~al.} 2008{\natexlab{a}},
  \aap, 486, L39

\bibitem[{{Fontaine} {et~al.}(2008{\natexlab{b}}){Fontaine}, {Chayer},
  {Oliveira}, {Wesemael}, \& {Fontaine}}]{fontm08}
{Fontaine}, M., {Chayer}, P., {Oliveira}, C.~M., {Wesemael}, F., \& {Fontaine},
  G. 2008{\natexlab{b}}, \apj, 678, 394

\bibitem[{{Geier}(2013)}]{geier13}
{Geier}, S. 2013, \aap, 549, A110

\bibitem[{{Gianninas} {et~al.}(2010){Gianninas}, {Bergeron}, {Dupuis}, \&
  {Ruiz}}]{gia10}
{Gianninas}, A., {Bergeron}, P., {Dupuis}, J., \& {Ruiz}, M.~T. 2010, \apj,
  720, 581

\bibitem[{{Harris}(1996)}]{har96}
{Harris}, W.~E. 1996, \aj, 112, 1487

\bibitem[{{Heber}(2008)}]{heb08}
{Heber}, U. 2008, \memsai, 79, 375

\bibitem[{{Heber}(2016)}]{heb16}
{Heber}, U. 2016, \pasp, 128, 082001

\bibitem[{{Hu} {et~al.}(2011){Hu}, {Tout}, {Glebbeek}, \& {Dupret}}]{hu11}
{Hu}, H., {Tout}, C.~A., {Glebbeek}, E., \& {Dupret}, M.-A. 2011, \mnras, 418,
  195

\bibitem[{{Johnson} {et~al.}(2014){Johnson}, {Green}, {Wallace}, {O'Malley},
  {Amaya}, {Biddle}, \& {Fontaine}}]{john14}
{Johnson}, C., {Green}, E., {Wallace}, S., {et~al.} 2014, in Astronomical
  Society of the Pacific Conference Series, Vol. 481, 6th Meeting on Hot
  Subdwarf Stars and Related Objects, ed. V.~{van Grootel}, E.~{Green},
  G.~{Fontaine}, \& S.~{Charpinet}, 153

\bibitem[{{Kilkenny} {et~al.}(1997){Kilkenny}, {Koen}, {O'Donoghue}, \&
  {Stobie}}]{kil97}
{Kilkenny}, D., {Koen}, C., {O'Donoghue}, D., \& {Stobie}, R.~S. 1997, \mnras,
  285, 640

\bibitem[{{Koen} {et~al.}(1997){Koen}, {Kilkenny}, {O'Donoghue}, {van Wyk}, \&
  {Stobie}}]{koen97}
{Koen}, C., {Kilkenny}, D., {O'Donoghue}, D., {van Wyk}, F., \& {Stobie}, R.~S.
  1997, \mnras, 285, 645

\bibitem[{{Lanz} \& {Hubeny}(2003)}]{lanz03}
{Lanz}, T. \& {Hubeny}, I. 2003, \apjs, 146, 417

\bibitem[{{Lanz} \& {Hubeny}(2007)}]{lanz07}
{Lanz}, T. \& {Hubeny}, I. 2007, \apjs, 169, 83

\bibitem[{{Latour} {et~al.}(2016){Latour}, {Chayer}, {Green}, \&
  {Fontaine}}]{lat16}
{Latour}, M., {Chayer}, P., {Green}, E.~M., \& {Fontaine}, G. 2016, ArXiv
  e-prints 1610.01306]

\bibitem[{{Latour} {et~al.}(2011){Latour}, {Fontaine}, {Brassard}, {Green},
  {Chayer}, \& {Randall}}]{lat11}
{Latour}, M., {Fontaine}, G., {Brassard}, P., {et~al.} 2011, \apj, 733, 100

\bibitem[{{Latour} {et~al.}(2015){Latour}, {Fontaine}, {Green}, \&
  {Brassard}}]{lat15}
{Latour}, M., {Fontaine}, G., {Green}, E.~M., \& {Brassard}, P. 2015, \aap,
  579, A39

\bibitem[{{Latour} {et~al.}(2014){Latour}, {Randall}, {Fontaine}, {Bono},
  {Calamida}, \& {Brassard}}]{lat14}
{Latour}, M., {Randall}, S.~K., {Fontaine}, G., {et~al.} 2014, \apj, 795, 106

\bibitem[{{Moehler} {et~al.}(2011){Moehler}, {Dreizler}, {Lanz}, {Bono},
  {Sweigart}, {Calamida}, \& {Nonino}}]{moe11}
{Moehler}, S., {Dreizler}, S., {Lanz}, T., {et~al.} 2011, \aap, 526, A136

\bibitem[{{Moni Bidin} {et~al.}(2012){Moni Bidin}, {Villanova}, {Piotto},
  {Moehler}, {Cassisi}, \& {Momany}}]{moni2012}
{Moni Bidin}, C., {Villanova}, S., {Piotto}, G., {et~al.} 2012, \aap, 547, A109

\bibitem[{{Moni Bidin} {et~al.}(2011){Moni Bidin}, {Villanova}, {Piotto},
  {Moehler}, \& {D'Antona}}]{moni2011}
{Moni Bidin}, C., {Villanova}, S., {Piotto}, G., {Moehler}, S., \& {D'Antona},
  F. 2011, \apjl, 738, L10

\bibitem[{{Napiwotzki}(1993)}]{nap93}
{Napiwotzki}, R. 1993, Acta Astronomica, 43, 343

\bibitem[{{{\O}stensen} {et~al.}(2010){{\O}stensen}, {Oreiro}, {Solheim},
  {Heber}, {Silvotti}, {Gonz{\'a}lez-P{\'e}rez}, {Ulla}, {P{\'e}rez
  Hern{\'a}ndez}, {Rodr{\'{\i}}guez-L{\'o}pez}, \& {Telting}}]{ost10}
{{\O}stensen}, R.~H., {Oreiro}, R., {Solheim}, J.-E., {et~al.} 2010, \aap, 513,
  A6

\bibitem[{{Randall} {et~al.}(2009){Randall}, {Calamida}, \& {Bono}}]{ran09}
{Randall}, S.~K., {Calamida}, A., \& {Bono}, G. 2009, \aap, 494, 1053

\bibitem[{{Randall} {et~al.}(2011){Randall}, {Calamida}, {Fontaine}, {Bono}, \&
  {Brassard}}]{ran11}
{Randall}, S.~K., {Calamida}, A., {Fontaine}, G., {Bono}, G., \& {Brassard}, P.
  2011, \apjl, 737, L27

\bibitem[{{Randall} {et~al.}(2016){Randall}, {Calamida}, {Fontaine}, {Monelli},
  {Bono}, {Alonso}, {Van Grootel}, {Brassard}, {Chayer}, {Catelan},
  {Littlefair}, {Dhillon}, \& {Marsh}}]{ran16}
{Randall}, S.~K., {Calamida}, A., {Fontaine}, G., {et~al.} 2016, \aap, 589, A1

\bibitem[{{Randall} {et~al.}(2005){Randall}, {Fontaine}, {Brassard}, \&
  {Bergeron}}]{ran05}
{Randall}, S.~K., {Fontaine}, G., {Brassard}, P., \& {Bergeron}, P. 2005,
  \apjs, 161, 456

\bibitem[{{Rauch} {et~al.}(2014){Rauch}, {Rudkowski}, {Kampka}, {Werner},
  {Kruk}, \& {Moehler}}]{rauch14}
{Rauch}, T., {Rudkowski}, A., {Kampka}, D., {et~al.} 2014, \aap, 566, A3

\bibitem[{{Rauch} {et~al.}(2010){Rauch}, {Werner}, \& {Kruk}}]{rauch10}
{Rauch}, T., {Werner}, K., \& {Kruk}, J.~W. 2010, \apss, 329, 133

\bibitem[{{Rauch} {et~al.}(2007){Rauch}, {Ziegler}, {Werner}, {Kruk},
  {Oliveira}, {Vande Putte}, {Mignani}, \& {Kerber}}]{rauch07}
{Rauch}, T., {Ziegler}, M., {Werner}, K., {et~al.} 2007, \aap, 470, 317

\bibitem[{{Rodr{\'{\i}}guez-L{\'o}pez}
  {et~al.}(2007){Rodr{\'{\i}}guez-L{\'o}pez}, {Ulla}, \& {Garrido}}]{rod07}
{Rodr{\'{\i}}guez-L{\'o}pez}, C., {Ulla}, A., \& {Garrido}, R. 2007, \mnras,
  379, 1123

\bibitem[{{Sandhaus} {et~al.}(2016){Sandhaus}, {Debes}, {Ely}, {Hines}, \&
  {Bourque}}]{sand16}
{Sandhaus}, P.~H., {Debes}, J.~H., {Ely}, J., {Hines}, D.~C., \& {Bourque}, M.
  2016, \apj, 823, 49

\bibitem[{{Seaton}(1979)}]{sea79}
{Seaton}, M.~J. 1979, \mnras, 187, 73P

\bibitem[{{Woudt} {et~al.}(2006){Woudt}, {Kilkenny}, {Zietsman}, {Warner},
  {Loaring}, {Copley}, {Kniazev}, {V{\"a}is{\"a}nen}, {Still}, {Stobie},
  {Burgh}, {Nordsieck}, {Percival}, {O'Donoghue}, \& {Buckley}}]{wou06}
{Woudt}, P.~A., {Kilkenny}, D., {Zietsman}, E., {et~al.} 2006, \mnras, 371,
  1497

\end{thebibliography}

\end{document}